\documentclass[final,3p,twocolumn]{elsarticle}

\usepackage{natbib}
\usepackage{hyperref}
\usepackage{xcolor}

\journal{Journal of Crystal Growth}

\begin{document}

	\begin{frontmatter}
		
		\title{Growth of layered Lu$ _2 $Fe$ _3 $O$ _7 $ and Lu$ _3 $Fe$ _4 $O$ _{10} $  single crystals exhibiting long-range charge order via the optical floating-zone method}

		\author[rvt]{S.S.~Hammouda\corref{cor1}}
		\ead{s.hammouda@fz-juelich.com}
		
		\author[rvt]{M. Angst}

		\address[rvt]{J\"{u}lich Centre for Neutron Science JCNS and Peter Grünberg Institut PGI, JARA-FIT, Forschungszentrum J\"{u}lich GmbH, 52425 Julich, Germany}
		
		\cortext[cor1]{Corresponding author}

		\begin{abstract}
We report the controlled growth of single crystals of intercalated layered Lu$ _{1+n} $Fe$ _{2+n} $O$ _{4+3n-\delta} $ ($ n $=1,2) with different oxygen stoichiometries ${\delta}$. For the first time crystals sufficiently stoichiometric to exhibit superstructure reflections in X-ray diffraction  attributable to charge ordering were obtained. The estimated correlation lengths tend to be smaller than for not intercalated LuFe$ _2 $O$ _4$. For Lu$ _2 $Fe$ _3 $O$ _7 $, two different superstructures were observed, one an incommensurate zigzag pattern similar to previous observations by electron diffraction, the other an apparently commensurate pattern with ($\frac{1}{3}\frac{1}{3}0$) propagation. Implications for the possible charge order in the bilayers are discussed. Magnetization measurements suggest reduced magnetic correlations and the absence of an antiferromagnetic phase.
			
		\end{abstract}
		
		
	\end{frontmatter}
	
	\section{Introduction}
	\label{}
	
Rare earth ferrites $R$Fe$_2$O$_4$ have attracted a lot of attention as proposed multiferroics. In particular, LuFe$ _2 $O$ _4$ was considered a clear example of ferroelectricity from charge ordering (CO) of Fe$ {^{2+}} $ and Fe$ {^{3+}} $ in the Fe-O bilayers \cite{Ikeda2005}, though recently this was contradicted \cite{ISI:000303384400013,Niermann2012,Ruff2012,Lafuerza2013}.  Rare earth substitutions tune the relevant interactions within the Fe-O bilayers \cite{Angst2013} resulting in a similar CO for $R$ = Yb, which has almost the same ion size as Lu \cite{williamson2017growth,Yb} but a dramatically different CO for the larger Y \cite{mueller2015stoichiometric,Y}. Another way to tune the CO is to focus on the interactions between different bilayers. This can be achieved by intercalating single Fe-O layers, increasing the distance bewtwenn the bilayers, which would reduce the likelihood of "charged bilayers" \cite{ISI:000303384400013} and thus make a ferroelectric CO more likely. That such intercalations of rare earth ferrites exist has been known since the 1970s \cite{kimizuka1974new,kimizuka1976new,matsui1979structure,matsui1980extra}, though few physical properties have been reported \cite{tanaka1983magnetic,iida1992magnetic,tanaka1993mossbauer,tanaka1994mossbauer,qin2009suppression,yang2010electronic}.

\par
\begin{figure}
 	\centering
 	\includegraphics[width={0.42\textwidth}]{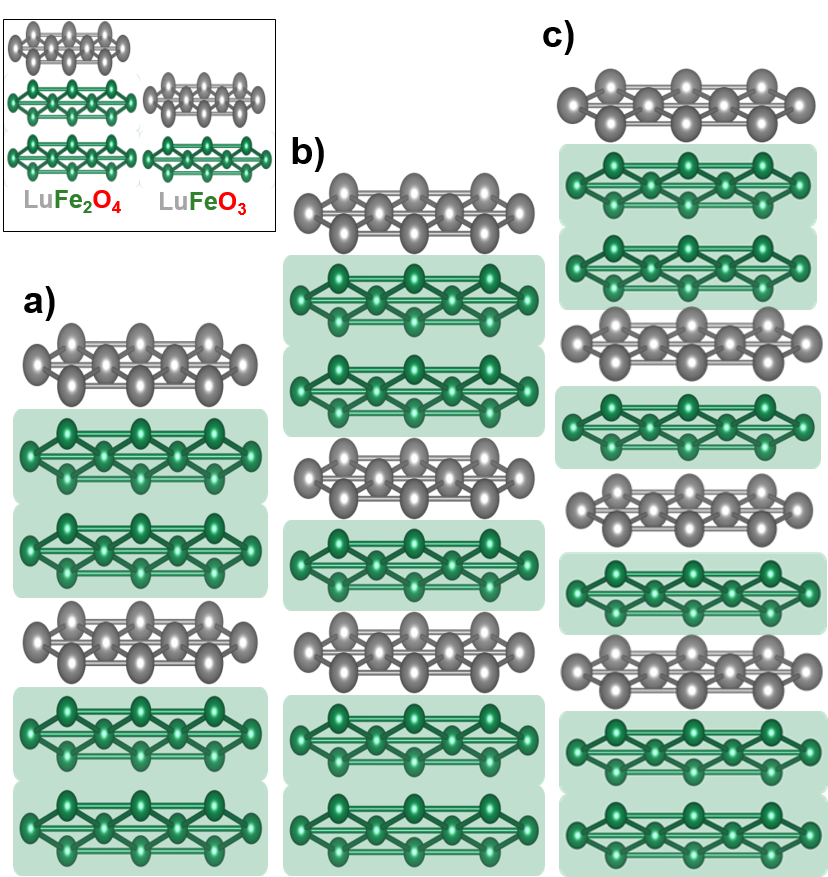}
 	\caption{Sketch of the layer stacking of a) LuFe$ _2 $O$ _4 $, b) Lu$ _2 $Fe$ _3 $O$ _7 $ containing one LuFeO$ _3 $ block, c) Lu$ _3 $Fe$ _4 $O$ _{10} $ containing two LuFeO$ _3 $ blocks (Oxygen ions are omitted).\label{Fig1}}
 \end{figure}

 In intercalated rare earth ferrites $R$Fe$ _2 $O$ _4 $($R$FeO$ _3 $)$n$, ($R$FeO$ _3 $)$n$ blocks are inserted alternately between the Fe-O bilayers (see Fig.\ \ref{Fig1}), forming a series of compounds that crystallize alternatingly in rhombohedral ($R$\={3}$m$, $n$ even) and hexagonal ($P$$6_{3}$/$mmc$, $n$ odd) space groups as found for the Yb-compound in \cite{kimizuka1976new,matsui1979structure,matsui1980extra}. Each $R$FeO$ _3 $ block contains a mono-layer of Fe-O and a mono-layer of $R$-O. The iron ion in $R$$ _2 $Fe$ _3 $O$ _{7-\delta}$ ($n$ = 1) has an average valance of 2.67 for $\delta$ = 0. M\"{o}ssbauer studies \cite{tanaka1983magnetic,tanaka1994mossbauer,tanaka1993mossbauer} indicate that the Fe-O mono-layer in LuFeO$ _3 $ block contains only Fe$ {^{3+}} $ ions, while the bilayer contains Fe$ {^{2.5+}} $ as in LuFe$ _2 $O$ _4 $. For $n$$>$1 this is also likely the case.

 \par Thus, the CO in the bilayers of intercalated rare earth ferrites is expected to be very similar as the CO in not intercalated ones, with the intercalation serving as another knob to tune the concrete 3D arrangement. However, the more complex crystal structure makes the synthesis of high quality single crystals more difficult. This complication is added to the problem of ensuring the proper oxygen stoichiometry already noted for not intercalated rare earth ferrites, where it was found to be critical to the elucidation of the CO that is established \cite{ISI:000303384400013,williamson2017growth,Yb,mueller2015stoichiometric,Y}. For these reasons, information about the CO in intercalated compounds is very scarce. The only study \cite{iida1992magnetic} by single-crystal diffraction reports the observation of a diffuse rod along ($\frac{1}{3}\frac{1}{3}$$\ell$), which corresponds to typical observations in off-stoichiometric $R$Fe$_2$O$_4$ \cite{Angst2013} and indicates the absence of long-range order. The observation of superstructure spots has been reported only from electron diffraction on small grains of polycrystalline Lu$ _2 $Fe$ _3 $O$ _7 $ \cite{qin2009suppression,yang2010electronic}. These spots form an incommensurate zig-zag pattern around the ($\frac{1}{3}\frac{1}{3}$$\ell$) line, which is consistent with a similar CO as in LuFe$ _2 $O$ _4$. However, electron diffraction is generally not suited to deduce the concrete CO pattern in real space. For this purpose, x-ray diffraction on sufficiently stoichiometric single crystals is needed.
 
 \par Here, we report the floating-zone growth of single crystals of intercalated Lu$ _2 $Fe$ _3 $O$ _7 $ and Lu$ _3 $Fe$ _4 $O$ _{10} $ with different oxygen contents tuned by modifying the oxygen partial pressure during growth. Single-crystal diffraction reveals superstructure reflections for crystals of both these compounds, although the estimated correlation lengths are significantly lower than those we found previously for optimized non-intercalated rare earth ferrites \cite{Niermann2012,williamson2017growth}. Intriguingly, for Lu$ _2 $Fe$ _3 $O$ _7 $, we not only found specimens exhibiting the same zig-zag pattern as found earlier by electron diffraction\cite{qin2009suppression,yang2010electronic}, but also crystals exhibiting apparently commensurate CO with ($\frac{1}{3}\frac{1}{3}$0) propagation. In addition to x-ray diffraction, magnetization data is reported as well.

	\section{Single crystal growth}

Following the same method used in preparing many of the rare earth ferrites (e.g. \cite{williamson2017growth,mueller2015stoichiometric}), powdered Lu$ _2 $O$ _3$ (99.9\%) and Fe$ _2 $O$ _3 $ (99.99\%) was mixed in stoichiometric quantities with respect to the metal ions. To get a homogeneous fine mixture, it was ground by ball milling. Pelleting the powder was necessary to ensure the reaction to be completed and avoid the appearance of white color identified as due to an impurity of Lu$ _2 $O$ _3 $. Afterward, the pellet was calcined in a tube furnace under controlled oxygen partial pressure using varying mixtures of flows of CO$_2$ and Ar(96\%):H$ _2 $(4\%) at 1250 $^\circ$C, for 40 hours. The oxygen partial pressure  resulting from using different gas ratios determines phase stability and oxygen stoichiometry \cite{Sekine197649}. Powder X-ray diffraction using a Huber Guinier D670 diffractometer (Cu-K${\alpha}$ radiation) was done for each prepared pellet calcined at specific CO$ _2 $-H$ _2$(4\%) gas flow to check the phase purity. Lu$ _2 $Fe$_3$O$ _{7-\delta}$ is found as a pure stable phase in a region with gas flows varying between 23-39 ml/min. CO$ _2 $ and 30 ml/min. Ar(96\%):H$_2$(4\%) (see supplementary material, Fig.\ S1). In contrast to LuFe$ _2 $O$_{4-\delta}$, the stoichiometry range for Lu$ _2 $Fe$ _3 $O$_{7-\delta}$ is wider according to \cite{Sekine197649}, with ${\delta}$ ranging from 0 to 0.104. Moreover, no region of surplus oxygen ($\delta$ $<$ 0) was reported in \cite{Sekine197649}, suggesting that the most stoichiometric compound will be near the upper phase stability range with respect to the oxygen partial pressure.

\par The raw ground powder was compressed using a hydraulic press to form rods of 5-6 cm in length then sintered in a flow of 27 ml/min. CO$ _2 $ and 30 ml/min. Ar(96\%):H$ _2 $(4\%) to maintain the phase purity. The floating zone method was used for crystal growth employing a four-mirror furnace FZ-T-10000-H-VI-VP0. This method was used successfully by \cite{iida1992magnetic} to prepare Lu$ _2 $Fe$ _3 $O$ _7$ single crystals, but without optimization for the oxygen stoichiometry. In the process of our crystal growth, we used a growth speed of 1-1.1 mm/hour, a rotation speed of 20 (16) rpm for the upper (lower) shaft and a gas flow of varying CO$ _2 $/CO ratio to tune the oxygen partial pressure during the growth. Fine tuning of the gas ratio was previously used to grow high-quality crystals of LuFe$ _2 $O$ _4 $ \cite{Christianson2008}, YFe$ _2 $O$ _4$ \cite{mueller2015stoichiometric} and YbFe$ _2 $O$ _4$ \cite{williamson2017growth}. However, stabilizing the molten zone was more difficult compared to LuFe$ _2 $O$ _4 $, which might be due to the complex layered structure and no stoichiometric single crystals were obtained, that are large enough for  e.g. neutron diffraction. The grown boule has length of 8 mm. 

\par
\begin{figure}
	\centering
	\includegraphics[width={0.45\textwidth}]{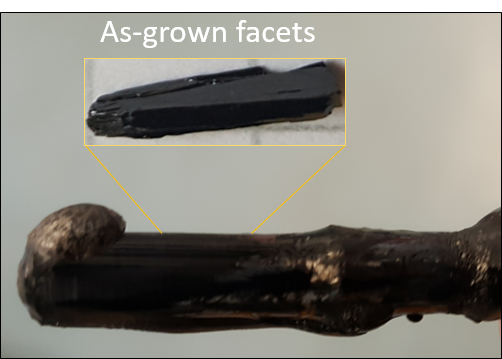}
	\caption{Crystal boule grown in gas flow of CO$_2$/ CO = 33 and (001) facets \label{Fig2}}
\end{figure}

 In analogy to LuFe$ _2 $O$ _4 $ \cite{joostphd}, the obtained crystals tend to cleave along the layers. Facets are formed because of the anisotropic distribution of the growth velocities, here in particular (001) facets are formed \cite{levy1977single,Shindo1976}. Fig.\ \ref{Fig2} shows the grown boule of Lu$ _2 $Fe$ _3 $O$ _7 $ in gas flow of CO$ _2 $/CO = 33 and a cleaved facet along the layer. Based on trial-and-error, many attempts with different gas ratios have been made to optimize the stoichiometry. In order to analyze the grown rod for each attempt, it was crushed, and the desired crystals were isolated by hand under the microscope.

	\section{Powder X-ray diffraction} 
	
\begin{figure}
	\centering
	\includegraphics[width={0.47\textwidth}]{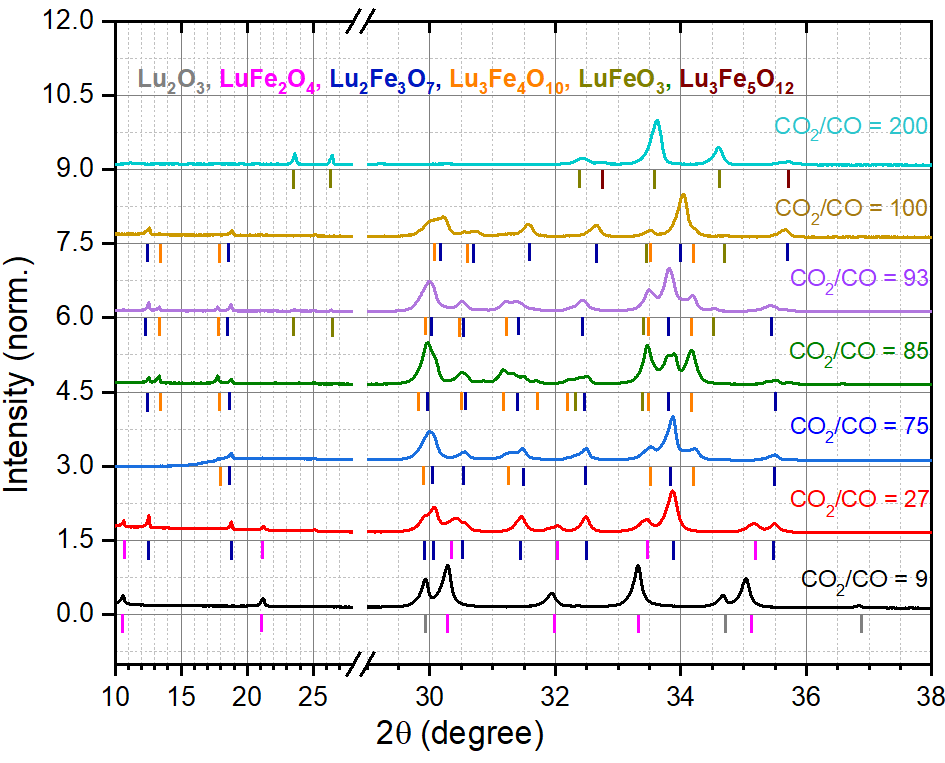}
	\caption{Powder diffractograms of powdered single crystal Lu$ _2 $Fe$ _3 $O$ _7 $ grown with different gas flow of CO$ _2$/CO, the present phases in each diffractogram marked with small lines. \label{Fig3}}
\end{figure}

For optimizing the synthesis conditions based on the presence of foreign phases, regions of the grown boule containing several crystals and potentially polycrystalline material from each growth attempt were ground and checked by powder X-ray diffraction at room temperature. Fig.\ \ref{Fig3} shows the  corresponding powder diffractograms   of parts of the grown boules for a few selected CO$_2 $/CO ratios. Starting with the lowest gas ratio CO$ _2 $/CO = 9 leads to the formation of LuFe$_2 $O$ _4 $ as the main phase rather than Lu$ _2 $Fe$ _3 $O$ _7$, and some impurity of Lu$ _2 $O$ _3 $ indicating the very low oxygen partial pressure following the phase diagram of \cite{Sekine197649}. In contrast, using a very high gas ratio of 200 leads to the growth of LuFeO$ _3 $ as main phase and some impurities of Lu$ _3 $Fe$ _5$O$ _{12} $ indicating the very high oxygen partial pressure as in \cite{Sekine197649}.

\par
\begin{figure}
 	\centering
 	\includegraphics[width={0.47\textwidth}]{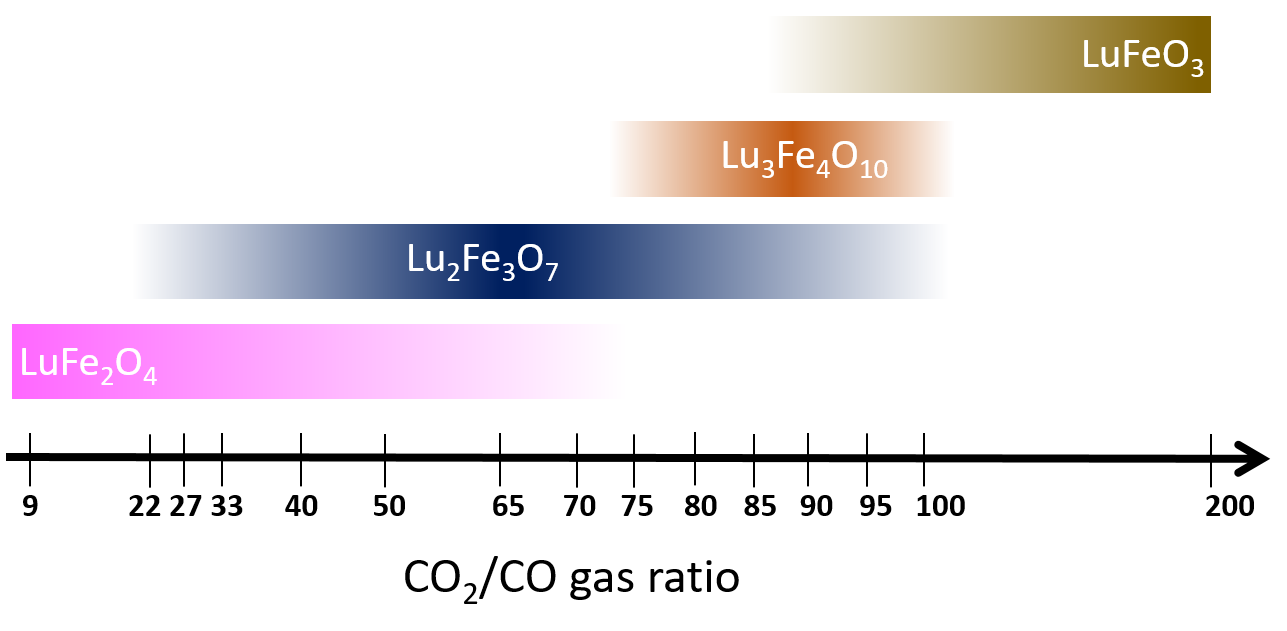}
 	\caption{Main phases obtained for some of the crystal growth attempts at different gas flow of CO$ _2 $/CO. The ticks at the bottom indicate the concrete gas ratios used in the various crystal growths. \label{Fig4}}
 \end{figure}

 The target phase Lu$ _2 $Fe$ _3 $O$ _7$ was observed in the range of CO$ _2 $/CO from 22 to 100, but it was never observed as the only phase, in contrast to our synthesized polycrystalline samples, see Fig.\ \ref{Fig4}. For growths in the range CO$ _2 $/CO = 75-80, neither LuFe$ _2 $O$ _4$ nor LuFeO$ _3 $ were present. However, new peaks which index to the second intercalation compound Lu$ _3 $Fe$ _4 $O$ _{10} $ are present. Additional peaks indexing to LuFeO$ _3$ are sometimes observed in the range of CO$ _2 $/CO = 80-100 besides those. Moreover, the powderized material was attracted by magnet suggesting the presence of a small phase fraction of Magnetite as well, which was not noticeable in the diffractogram because magnetite is weakly diffracting. The presence of both LuFeO$ _3 $ and Fe$_3$O$_4$ is an indication that we are around the upper stability limit of the Lu$ _2 $Fe$ _3 $O$ _7$, therefore in the region of most stoichiometric Lu$ _2 $Fe$ _3 $O$ _7$ according to \cite{Sekine197649}. Lu$ _3 $Fe$ _4 $O$ _{10} $ is also present at this upper stability limit.

	\section{Single-crystal X-ray diffraction: probing long range charge order (CO)}
	
	\begin{figure}
		\centering
		\includegraphics[width={0.48\textwidth}]{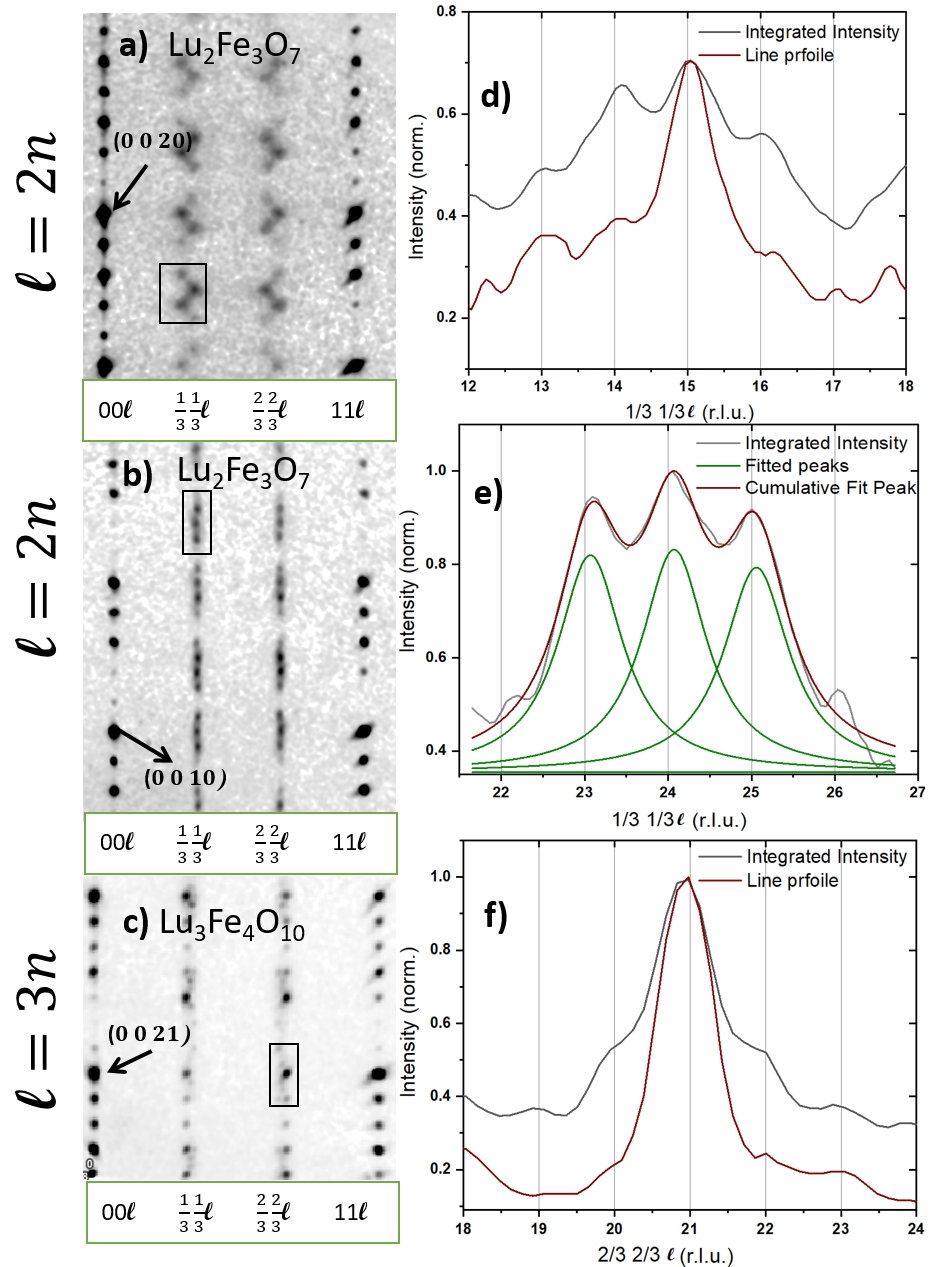}
		\caption{ Precession images of the (hh$\ell$) reciprocal space plane (with the intensity in a small region perpendicular to (hh$\ell$) integrated) for (a-c) and cuts along $\ell$ (d-f) for (a,d) Lu$ _2 $Fe$_3$O$ _{7} $ (b,e) Lu$ _2 $Fe$ _3 $O$ _7$ (c,f) Lu$ _3 $Fe$ _4 $O$ _{10} $. All crystals were grown with CO$ _2 $/CO = 85 and data was collected at room temperature. \label{Fig5}}
	\end{figure}

 The charge order CO was investigated at room temperature using a Rigaku Supernova diffractometer employing Mo-K${\alpha}$ radiation, as already used to determine the CO of LuFe$ _2 $O$ _4 $ \cite{ISI:000303384400013}, YbFe$ _2 $O$ _4 $ \cite{williamson2017growth,Yb} and YFe$ _2 $O$ _4 $ \cite{mueller2015stoichiometric,Y}.  Many crystals that were prepared with gas ratio 80-100 in which the most stoichiometric Lu$ _2 $Fe$ _3 $O$ _7 $ is expected, c.f.\ (Sec.\ 2 and 3) were checked. Crystals of Lu$ _2 $Fe$ _3 $O$ _7 $ or Lu$ _3 $Fe$ _4 $O$ _{10} $ were found, but, unlike reported in \cite{matsui1980extra} no instances of intergrowths of both phases in the same crystal were found.  
 
  \begin{figure}
 	\centering
 	\includegraphics[width={0.5\textwidth}]{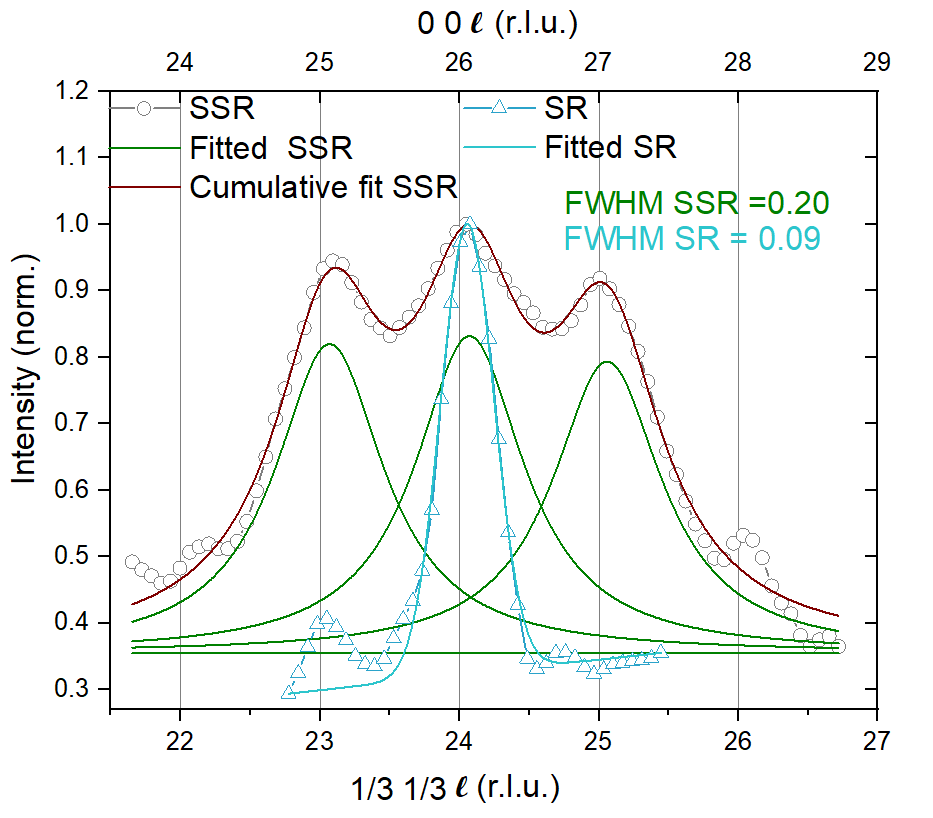}
 	\caption{Intensity integrated perpendicular to $\ell$ of: (0 0 26) structural reflection (SR), fitted SR intensity and super structural reflection (SSR) ($\frac{1}{3}\frac{1}{3}$$\ell$ = 23-25) out of plane at room temperature. Full widths at half maximum (FWHM) are given in Angstrom units. \label{Fig6}}
 \end{figure}

 \par Regarding Lu$ _2 $Fe$ _3 $O$ _7$, three different types of diffraction results were obtained: off-stoichiometric crystals showing a zigzag diffuse scattering along ($\frac{1}{3}\frac{1}{3}$$\ell$) in addition to Bragg reflections from the $P$6$_{3}$/$mmc$ basic crystal structure (see supplementary material, Fig.\ S2), the second type of crystals exhibits superstructure reflections also with zigzag pattern, as can be seen in the projection of the reciprocal $hh\ell$ plane (Fig.\ \ref{Fig5}a). The superstructure reflections can be indexed with incommensurate propagation vector ($\frac{1}{3}$$-$$\tau$$,\frac{1}{3}$$-$$\tau$,0) and symmetry-equivalent, with values of $\tau$ up to 0.025 ($\tau$ $\sim$ 0.022 for Fig.\ \ref{Fig5}a) . This type of incommensurate pattern had been reported for polycrystalline Lu$ _2 $Fe$ _3 $O$ _7 $ using electron diffraction in \cite{yang2010electronic}. The intensity integrated in $h$$h$-direction around $h$$h$ = 1/3 vs $\ell$ is shown in Fig.\ \ref{Fig5}d, also shown is a line profile through the center of one of the peaks (red line). 
 
 \par The third type exhibits a commensurate ($\tau$ = 0 within experimental resolution) superstructure with ($\frac{1}{3}\frac{1}{3}$0) propagation (Fig.\ \ref{Fig5}b). The intensity integrated in $h$$h$-direction around $h$$h$ = 1/3 vs $\ell$, the fitted individual peaks (green line) and the cumulative fit peak (grey) are shown in Fig.\ \ref{Fig5}e. Such a commensurate CO was not observed before in LuFe$ _2 $O$ _4$ or Lu$ _2 $Fe$ _3 $O$ _7 $, but there is one report where such a commensurate pattern was found in YbFe$ _2 $O$ _4$ \cite{nagata2018modulation}. 
 
 \par  Lu$ _3 $Fe$ _4 $O$ _{10} $ crystals also exhibit both diffuse scattering (see supplementary material, Fig.\ S3) and superstructure reflections with a zigzag pattern. The superstructure reflections can be indexed with incommensurate propagation vector ($\frac{1}{3}$$-$$\tau$$,\frac{1}{3}$$-$$\tau$,0), with values of $\tau$ up to 0.019, see (Fig.\ \ref{Fig5}c, $\tau$ = 0.012). The intensity integrated and line profile along ($\frac{2}{3}\frac{2}{3}$ $\ell$) are also shown for this compound in Fig.\ \ref{Fig5}f.

 \par To estimate the out-of-plane correlation lengths at room temperature, a comparison of the peak width of the super structural reflection (SSR) ($\frac{1}{3}\frac{1}{3}$ 24) and the structural reflection (SR) (0 0 26) is shown in Fig.\ \ref{Fig6}. Subtracting the width of the SSR from SR in order to approximately correct for the effect of instrumental resolution and mosaicity, provides an estimated correlation length of 27 {\AA} for the incommensurate and 19 {\AA} for the commensurate CO. The correlation length for Lu$ _3 $Fe$ _4 $O$ _{10} $ is calculated in the same manner to be 49 {\AA}. The correlation lengths for both compounds are smaller than the correlation length reported in LuFe$ _2 $O$ _4 $ (75 {\AA} \cite{Wen2009}), and also smaller than we observed in YFe$ _2 $O$ _4 $ (550 {\AA} \cite{mueller2015stoichiometric}). The shorter correlation lengths observed in the intercalated compounds are likely due to the larger separation of the bilayers in which the CO takes place. Nevertheless, the correlations are sufficient to deduce the CO pattern in principle.
    
\par Focusing on the commensurate CO in Lu$ _2 $Fe$ _3 $O$ _7 $, the superstructural reflections can be indexed by a propagation vector ($\frac{1}{3}\frac{1}{3}$0), which leads to the same likely CO configurations as discussed for LuFe$ _2 $O$ _4 $ \cite{Angst2013} : either charged bilayers or polar bilayers stacked with the same or alternating polarizations  (note that because the Lu$ _2 $Fe$ _3 $O$ _7 $ unit cell contains two bilayers, an antipolar stacking corresponds to ($\frac{1}{3}\frac{1}{3}$0) propagation rather than ($\frac{1}{3}\frac{1}{3}\frac{3}{2}$) as in LuFe$ _2 $O$ _4 $). The same CO within a single bilayer as in LuFe$ _2 $O$ _4$ and YbFe$ _2 $O$ _4$ can indeed be expected, given that intralayer Fe-Fe distance to bilayer thickness (1.426) is very close to what is found for these two compounds (c.f.\ \cite{Angst2013}). Due to the different (hexagonal rather than rhombohedral) layer stacking, any CO with polar bilayers would imply a net polarization, however.

\par To decide which of these possible configurations is actually realized in Lu$ _2 $Fe$ _3 $O$ _7 $ requires a collection of a full data set of integrated intensities and structural refinement, as previously done for the non-intercalated compounds \cite{ISI:000303384400013,Yb,Y}. The shorter correlation lengths make this endeavor more difficult, as it contributes to significant peak overlap as can be seen in (Fig.\ \ref{Fig6}). This problem can be ameliorated by improving the experimental resolution, collecting data at a synchrotron beamline. Corresponding studies are in progress.

	\section{Magnetization measurements}
	
	\begin{figure}
	\centering
	\includegraphics[width={0.47\textwidth}]{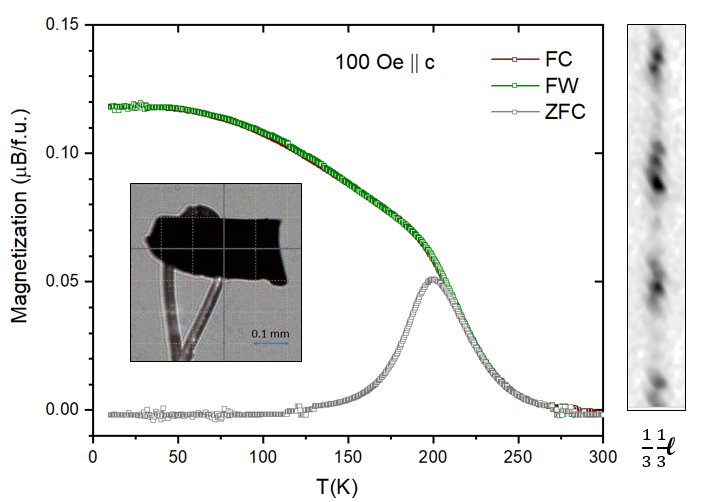}
	\caption{ Left: Magnetization measurements for stoichiometric single crystal Lu$ _2 $Fe$ _3 $O$ _7 $ measured during field cooling(FC), field warming (FW) and zero field cooling (ZFC), all measurements were done under an applied magnetic field of 100 Oe, measured with a rate 0.5 K/min. The inset shows the (0.4$\times$0.2$\times$0.05) mm$^3$ measured crystal mounted on holder for single crystal X-ray diffraction. Right: superstructural reflections along ($\frac{1}{3}\frac{1}{3}$$\ell$). \label{Fig7}}
\end{figure}	
	
We have seen the influence of stoichiometry on the appearance of CO in the Lu$ _2 $Fe$ _3 $O$ _7 $ crystals, and based on the results found for LuFe$ _2 $O$ _4 $ \cite{ISI:000299821100008}, YFe$ _2 $O$ _4 $ \cite{mueller2015stoichiometric} and YbFe$ _2 $O$ _4 $ \cite{williamson2017growth}, we expect a dependence of the magnetic properties on oxygen content as well. So far, magnetization measurements on a non-stoichiometric Lu$ _2 $Fe$ _3 $O$ _7 $ single crystal exhibiting 2D magnetic ordering has been reported in \cite{iida1992magnetic}. The Quantum Design Magnetic Property Measurement System (MPMS) was used to perform the magnetization measurements.

\par An example of a Lu$ _2 $Fe$ _3 $O$ _7 $ crystal grown in CO$ _2 $/CO = 85 is shown in the inset of Fig.\ \ref{Fig7}. This crystal exhibit superstructure reflections by single crystal X-ray diffraction (see Fig.\ \ref{Fig7} right) and was used for the magnetization measurements. The field was applied parallel to the $\vec{c}$-direction due to the strong magnetic anisotropy reported in \cite{iida1992magnetic}. Our crystal (see Fig.\ \ref{Fig7} left) reveals a ferrimganetic transition around 200 K in ZFC, with no indication for an antiferromagnetic phase as found in LuFe$ _2 $O$ _4 $ \cite{ISI:000303384400013}. Moreover, a large difference between ZFC and FC is noticeable indicating a glassy behavior without long range spin ordering. This suggests that our crystal exhibits a 3D CO but not 3D spin order (SO), indicating that the SO is more fragile. A similar observation was made before for some crystals of YbFe$ _2$O$ _4$ \cite{Yb}. No thermal hysteresis is present indicating that no first order transition took place. This single crystal shows a similar magnetic behavior as polycrsytalline Lu$ _2 $Fe$ _3 $O$ _7 $ in \cite{yang2010electronic}, but with lowering in the peak of the ZFC curve by $\sim$50 K.

\par As mentioned above, we expect the same CO as for LuFe$ _2 $O$ _4 $ or YbFe$ _2 $O$ _4 $ to be realized in a single bilayer in Lu$ _2 $Fe$ _3 $O$ _7 $, and because of the strong spin-charge coupling \cite{ISI:000303384400013,Angst2013,Yb} the same SO may be is expected as well. In contrast to LuFe$ _2 $O$ _4 $ and YbFe$ _2 $O$ _4 $, where both competing phases of antiferromgantic and ferrimagnetic are present that differ only in the stacking of the bilayer net magnetizations \cite{ISI:000299821100008, Yb}, our result suggests a preference for the ferrimagnetic phase to be stabilized in the Lu$ _2 $Fe$ _3 $O$ _7 $ as a result of the modified magnetic interactions between neighboring bilayers.

	\section{Conclusion and outlook}
Based on our interest of investigating the CO in the intercalated compound, we succeeded in growing single crystals of Lu$ _2 $Fe$ _3 $O$ _7 $, but also Lu$ _3 $Fe$ _4 $O$ _{10} $, which are sufficiently stoichiometric to exhibit for the first time superstructure reflections indicating the long range charge order. The estimated correlation lengths are smaller than the one for LuFe$ _2 $O$ _4 $. The availability of these crystals open the door to continue to the refinement of CO and answering the question of ferroelectricity which is in progress.  	
	\label{}
	
	\section{Acknowledgments}
We gratefully acknowledge J\"{o}rg Per{\ss}on for assistance during crystal growth. This work was funded in part by the Brain gain fund of Forschungszentrum J\"{u}lich GmbH.

\end{document}